# Telling the Early Story of Solar Energy Meteorology by Applying (Co-Citation) Reference Publication Year Spectroscopy

Thomas Scheidsteger[1] and Robin Haunschild[2]

[1] *t.scheidsteger@fkf.mpg.de*
Max Planck Institute for Solid State Research, Heisenbergstr. 1, D-70659 Stuttgart (Germany)

[2] *r.haunschild@fkf.mpg.de*
Max Planck Institute for Solid State Research, Heisenbergstr. 1, D-70659 Stuttgart (Germany)

**Abstract**
Studying the history of research fields by analyzing publication records and topical and/or keyword searches with a full Reference Publication Year Spectroscopy (RPYS) has been introduced as a powerful tool to identify the corresponding root publications. However, for a rather new and interdisciplinary research field like Solar Energy Meteorology (SEM), this method is not feasible to get a reasonably exhaustive publication set. Therefore we apply its variant RPYS-CO to all publications co-cited with one highly important marker paper, using the CRExplorer for plotting and inspecting the spectrogram of the number of cited references. Examining its peaks and their main contributing publications, we get a list of seminal papers, which are able to adequately tell us the story of SEM up to the 1990s. Generally, we recommend this method to gain valuable insights in (new) research fields.

**Introduction**
Solar Energy Meteorology (SEM) studies how solar radiation can be utilized for solar energy conversion to provide heat or electricity and how the performance of these conversion processes is affected by meteorological influences. This is largely the question of its availability in time (e.g., time of day, year) and space (e.g., geographical location, angular orientation). The quality of radiation then matters when different devices are used: Concentrating devices need *direct* radiation to work whereas non-concentrating photovoltaics, i.e., solar cells, can also utilize the *diffuse* fraction of sunlight, usually scattered in the atmosphere. So, the main fields of investigations are (i) measurements and their evaluation over different time scales and (ii) modeling of radiation and its components, depending on physical (e.g., solar constant, equation of time), geometrical (e.g., position of the sun, orientation of the converter) and meteorological (e.g., cloud coverage, aerosol concentration) parameters. Both fields also involve a lot of statistical treatment.

With the availability of geostationary weather satellites, for example the European METEOSAT, methods based on satellite data have been developed in several research groups since the 1980s. One of the authors has been a member of Oldenburg University's research group on Energy Meteorology for some time, so it came naturally to investigate this field of research with bibliometric methods in order to identify seminal and landmark papers, that lead up to the state of the art at the beginning of the satellite era in SEM.

Due to its intrinsic interdisciplinarity, potential search terms tend to have multiple meanings, which leads to answer sets from title or topical searches with low precision and/or recall. Therefore, we use the bibliometric method Reference Publications Year Spectroscopy (RPYS), introduced by Marx, Bornmann, Barth, and Leydesdorff (2014), but in a variant called RPYS-CO (for co-citation), where there is no need for an exhaustive paper set covering most of the research field. In RPYS-CO, the analyzed publication set is defined by at least one marker paper. All publications co-cited with this marker paper are included in the RPYS analysis. The RPYS has successfully been applied to identify the root publications of climate change research by Marx, Haunschild, Thor, and Bornmann (2017). In that case a set of more than 200,000 papers has been used. In a subsequent approach in the same paper, Marx, et al.

(2017) refined this large set to the greenhouse effect by keeping only cited references that are co-cited with Arrhenius (1896) and were able to retrieve the results of the RPYS on the full publication set regarding the greenhouse effect, but also lesser known works of relevance. They named this RPYS variant RPYS-CO. In another very recent study, (Haunschild & Marx, 2019) compare their own results of a RPYS on density functional theory, a very frequently applied method in computational chemistry (Haunschild, Barth, & Marx, 2016), with an RPYS-CO using *one single* seminal paper with a high citation count and find a striking similarity with the results of the analysis based on a search in controlled vocabulary.

Encouraged by these results, we set out here to investigate all publications co-cited with one highly-cited marker paper, a choice discussed with and corroborated by the long term leader of the Oldenburg group: *The interrelationship and characteristic distribution of direct, diffuse and total solar radiation* (Liu & Jordan, 1960). In order to indicate the importance of this marker paper we quote its complete abstract, emphasizing in italics all those concepts and terms that proved to be prevalent in SEM for its whole history:

"Based upon the data now available, this paper presents relationships permitting the determination on a *horizontal surface* of the instantaneous intensity of *diffuse radiation* on *clear days*, the *long term average hourly and daily sums* of diffuse radiation, and the daily sums of diffuse radiation for various categories of days of *differing degrees of cloudiness*. For these determinations, it is necessary to have, either from actual measurements or estimates, a knowledge of the *total (direct plus diffuse) radiation on a horizontal surface* – its *measurement is now regularly made* at 98 localities in the United States and Canada. For localities where only an *estimate of the long term average total radiation* is available, *relation-ships* presented in this paper can be utilized to determine the *statistical distribution of the daily total radiation at these localities*." (Liu & Jordan, 1960, abstract)

Satellite-based studies often view this paper as text book knowledge. Therefore it is affected by obliteration by incorporation (Cronin & Sugimoto, 2014) and rarely cited in this area of SEM, that gained traction in the late 1980s and early 1990s. The latter are consequently a natural end date for our study of cited references, co–cited with (Liu & Jordan, 1960). The then flourishing satellite-based publications could and should be the target of further investigations using other marker papers. On the other hand, due to the recency of the research field, we do not expect decisive contributions to SEM before 1900.

There are some studies with very different time frame, focus or methodology, e.g.: Du, Li, Brown, Peng, and Shuai (2014) analysed the solar energy literature from 1992 to 2011, but with no consideration of energy meteorology topics. A bibliometric analysis on solar power research between 1991 and 2010, again after the period of our study, has been performed by Dong, Xu, Luo, Cai, and Gao (2012) using terms as, e.g., "solar radiation" in a topical search in the WoS. Their goal was to identify research trends for the twenty-first century and not to explore historical roots. In the same vein Yang, Kleissl, Gueymard, Pedro, and Coimbra (2018) tried to identify key innovations for the future of research in "solar radiation and PV power forecasting", a field mainly emerging at the turn of the millennium. They based their work on the first 1000 hits of a keyword search in Google Scholar and applied machine learning and text mining methods to full texts in order to complement conventional topical reviews.

To the best of our knowledge, the present study is one of the first using the method RPYS-CO in order to identify seminal papers for a research field – thereby complementing qualitative knowledge of experts by a quantitative evaluation of the citation counts (i.e., the reference counts within the topic related literature). Using this method we were confident to find by this method those important contributors and their papers which tell the story of the emergence of solar energy meteorology from around 1900 up to the beginning of the 1990s. So we support the suggestion of (Haunschild & Marx, 2019) that this method can help researchers to explore their field of study - in a way complementary to a usual topical or keyword-based literature search.

**Method and data set**

As of 8 January 2019, Liu and Jordan (1960) had 1032 citing papers in the WoS until the end of 2018. One fourth of these papers (n=257, 25%) as well as the marker paper itself have been published in a single journal, *Solar Energy*. Their four most important WoS subject categories in the data set used in this study are *Energy Fuels* (n=673, 65%), *Green Sustainable Science Technology* (n=151, 15%), *Meteorology & Atmospheric Sciences* (n=131, 13 %), and *Thermodynamics* (n=114, 11%), thereby reflecting the multiple foci of SEM.

We downloaded the bibliographic data of the 1032 papers including 36,635 cited references (CRs) from the WoS (selecting "Save to Other File Formats" and "Other Reference Software") and imported them into the CRExplorer. (The JAVA based software can be downloaded for free from http://crexplorer.net and a comprehensive handbook explaining all functions is also available.) It provides a graphical display of the number of CRs (NCR) over the reference publication years (RPY) and a tabular presentation of the NCR of all CRs. In our case there were only single occurrences of CRs before 1900. After 1995, despite a sequence of steadily increasing peaks, no specific papers of main contribution (more than a share of 10% of the NCRs in the specific RPY) could be identified. Both facts confirm our choice of the time period to be analyzed.

Much of the processing can comprehensively and reproducibly be done by using the CRExplorer scripting language: With the script in Listing 1 we `import`ed the WoS file and got 8383 unique reference variants for the reference publication years 1900 to 1995. After that `clustering` and `merging` of equivalent CR variants was done with Levenshtein threshold 0.75 and taking volume and (starting) page number into account, thereby reducing the number of CR variants by 109. Then we `removed` all publications with only *one* citation, in order to reduce noise. In the end, we retained 1566 CRs. The results including the NCR and other indicators were `export`ed to CSV files for further inspection and plotting of the spectrogram, which can be done by using the R package BibPlots (see: https://cran.r-project.org/web/packages/BibPlots/index.html and https://tinyurl.com/y97bb54z).

```
importFile(file:"savedrecs_Liu_1960.txt",type:"WOS",
RPY:[1900,1995,false], PY:[1962,2018,false], maxCR:0 )
info()
cluster(threshold:0.75,volume:true,page:true,DOI:false)
merge()
info()
removeCR(N_CR: [0, 1])
info()
saveFile(file:"Liu1960.rpys.cre")
exportFile(file:"Liu_1960.rpys_CR.csv",type:"CSV_CR")
exportFile(file:"Liu_1960.rpys_GRAPH.csv",type:"CSV_GRAPH")
```
**Listing 1. CRExplorer script to perform RPYS on the WoS papers citing Liu and Jordan (1960)**

In the spectrogram, we looked for publication years with significantly higher NCR than other years, aided by the deviation of NCR from the 5-year-median of NCR (taking into account the two preceding and the two following years). For the papers that, by applying this methodology, seemed primarily responsible for the peaks a manual merging was done, if needed.

**Results**

Figure 1 shows the spectrogram of the RPYS-CO for the marker paper (Liu & Jordan, 1960) in terms of the NCR and their 5-year-median deviation for the whole analyzed time period, and Table 1 lists all publications, contributing substantially to the peaks of NCR and identified as relevant.

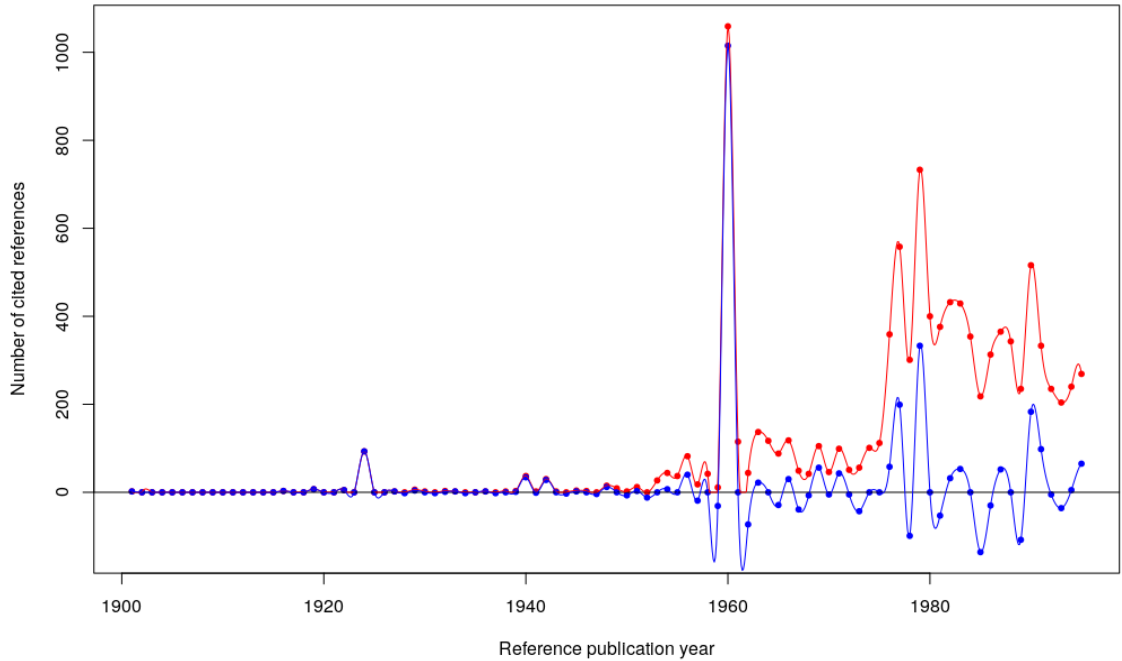

**Figure 1. Overall RPYS-CO graph for Liu and Jordan (1960) with NCR (red line) and 5-year-median deviation (blue line)**

**Table 1. RPYS-CO for Liu and Jordan (1960): important CRs from 1900 to 1995 with number of citations NCR and indicating, if manually merged during inspection of spectrogram**

| #CR | RPY | Cited Reference (Manually merged: *M*) | NCR |
| --- | --- | --- | --- |
| CR1 | 1919 | Kimball HH, 1919, Monthly Weather Review, V47, P769 | 7 |
| CR2 | 1922 | Angström A, 1922, Ark Mat Astron Fys, V17, P1 | 3 |
| CR3 | 1922 | Linke F, 1922, Beitr Phys Atmos, V10, P91 | 2 |
| CR4 | 1924 | Angström A, 1924, Quarterly Journal of the Royal Meteorological Society, V50, P121 *(M)* | 93 |
| CR5 | 1929 | Angström A, 1929, Geogr Annlr Stockhol, V11, P156 *(M)* | 6 |
| CR6 | 1940 | Prescott J, 1940, T Roy Soc South Aust, V64, P114 | 30 |
| CR7 | 1942 | Hottel HC, 1942, Transactions of the ASME, V64, P91 *(M)* | 23 |
| CR8 | 1945 | Haurwitz B, 1945, J Met, V2, P154 | 4 |
| CR9 | 1946 | Haurwitz B, 1946, J Met, V3, P123 | 3 |
| CR10 | 1948 | Haurwitz B, 1948, J Met, V5, P110 | 5 |
| CR11 | 1953 | Whillier A, 1953, Thesis, MIT Cambridge *(M)* | 17 |
| CR12 | 1954 | Black JN, 1954, Q J Roy Meteor Soc, V80, P231 | 28 |
| CR13 | 1955 | Hottel HC, 1955, T C Use Solar Energy, V2, P74 | 27 |
| CR14 | 1956 | Whillier A, 1956, Arch Meteorol Geophys U Bioklimatol Ser B, | 38 |

| | | V7, P197 *(M)* | |
|---|---|---|---|
| CR15 | 1958 | Glover J, 1958, Q J Roy Meteor Soc, V84, P172 | 18 |
| CR16 | 1960 | Liu BYH, 1960, Solar Energy, V4, P1 | 1031 |
| CR17 | 1963 | Liu BYH, 1963, Solar Energy, V7, P53 | 71 |
| CR18 | 1963 | Choudhury NKD, 1963, Solar Energy, V7, P44 | 37 |
| CR19 | 1964 | Page JK, 1964, P UN C New Sources E, V4, P378 *(M)* | 87 |
| CR20 | 1966 | Stanhill G, 1966, Solar Energy, V10, P96 | 29 |
| CR21 | 1966 | Robinson N, 1966, Solar Radiation *(M)* | 25 |
| CR22 | 1966 | Kasten F, 1966, Arch Meteorol Geop B, VB14, P206 | 11 |
| CR23 | 1969 | Cooper PI, 1969, Solar Energy, V12, P333 *(M)* | 42 |
| CR24 | 1969 | Kondratyev KY, 1969, Radiation in the Atmosphere | 23 |
| CR25 | 1971 | Spencer J, 1971, Search, V2, P172 *(M)* | 35 |
| CR26 | 1974 | Duffie JA, 1974, Solar Energy Thermal Processes *(M)* | 35 |
| CR27 | 1976 | Ruth DW, 1976, Solar Energy, V18, P153 | 68 |
| CR28 | 1976 | Hottel HC, 1976, Solar Energy, V18, P129 | 63 |
| CR29 | 1976 | Tuller SE, 1976, Solar Energy, V18, P259 | 46 |
| CR30 | 1976 | Hay JE, 1976, Atmosphere, V14, P278 | 34 |
| CR31 | 1977 | Orgill JF, 1977, Solar Energy, V19, P357 | 149 |
| CR32 | 1977 | Klein SA, 1977, Solar Energy, V19, P325 | 121 |
| CR33 | 1977 | Temps RC, 1977, Solar Energy, V19, P179 *(M)* | 46 |
| CR34 | 1979 | Collares-Pereira M, 1979, Solar Energy, V22, P155 | 238 |
| CR35 | 1979 | Klucher TM, 1979, Solar Energy, V23, P111 | 60 |
| CR36 | 1979 | Hay JE, 1979, Solar Energy, V23, P301 | 59 |
| CR37 | 1980 | Duffie JA, 1980, Solar engineering of thermal processes, 1$^{st}$ Ed. | 63 |
| CR38 | 1980 | Iqbal M, 1980, Solar Energy, V24, P491 | 46 |
| CR39 | 1981 | Bendt P, 1981, Solar Energy, V27, P1 | 66 |
| CR40 | 1982 | Erbs DG, 1982, Solar Energy, V28, P293 | 208 |
| CR41 | 1983 | Iqbal M, 1983, Introduction to Solar Radiation *(M)* | 117 |
| CR42 | 1987 | Skartveit A, 1987, Solar Energy, V38, P271 | 50 |
| CR43 | 1987 | Perez R, 1987, Solar Energy, V39, P221 | 42 |
| CR44 | 1988 | Suehrcke H, 1988, Solar Energy, V40, P413 | 36 |
| CR45 | 1988 | Graham VA, 1988, Solar Energy, V40, P83 *(M)* | 28 |
| CR46 | 1990 | Reindl DT, 1990, Solar Energy, V45, P1 | 116 |
| CR47 | 1990 | Reindl DT, 1990, Solar Energy, V45, P9 | 54 |
| CR48 | 1990 | Perez R, 1990, Solar Energy, V44, P271 | 49 |
| CR49 | 1991 | Duffie JA, 1991, Solar engineering of thermal processes, 2$^{nd}$ Ed. *(M)* | 83 |

The overall RPYS-CO picture can easily be divided by the maximum NCR per RPY into two periods with regard to the reference publication years, which we are going to discuss separately: the first one from 1915 leading to, but excluding, 1960, the publication year of the marker paper, containing peaks with at most NCR=100; the second one from 1960 to 1995, with peaks between NCR=100 and NCR=900 (apart from the marker paper itself).

*Time Period 1: 1915 to 1959*

Figure 2 shows the RPYS-CO spectrogram for the marker paper (Liu & Jordan, 1960) for the relevant time period before it was published.

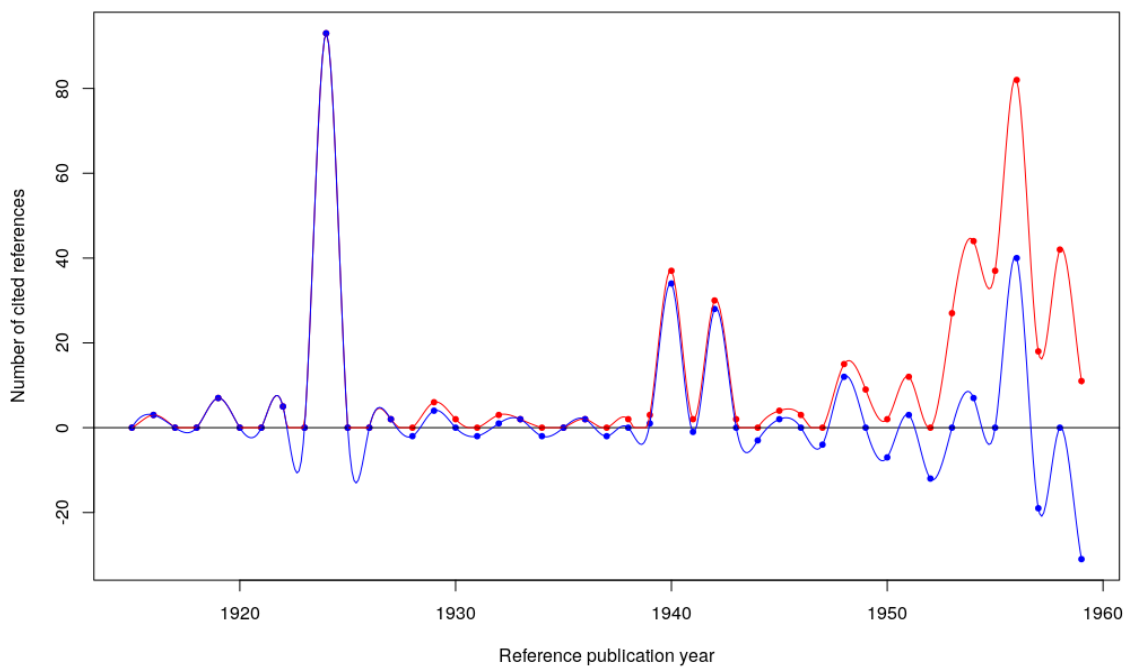

**Figure 2. RPYS-CO graph for Liu and Jordan (1960) and period 1 (1915 – 1959) with NCR (red line) and 5-year-median deviation (blue line).**

In this time period, we were able to identify 9 peaks with *relevant* papers for the following RPYs: 1919, 1922, 1924, 1929, 1940, 1942, 1945/46, 1948, and 1953-58. Because of the generally low NCR values in this time period, we did not want to lose reference variants of possibly relevant papers and therefore additionally looked at the CRs *before* the removal of only once referenced papers. But this did not reveal new relevant papers, instead it only confirmed the results from the reduced set.

Now we can follow the path of SEM by looking at the peak papers and drawing partially from explanations given in the citing papers. In this first period, two independent streams of research flew together: *meteorology* and *engineering*.

*Meteorologists* tried to develop a climatology of irradiance, emphasizing daily mean values, with no or little application to solar energy in mind. Solar irradiance varies deterministically with the sun's position on the sky dome and irregularly with changing cloudiness. The relation of the latter with sunshine has initially been measured by Kimball (CR1) and later subjected to statistical analysis by Angström (CR2, CR4, and CR5), leading to a linear relation between the duration of bright sunshine and average solar energy available on a horizontal surface at ground level, the so-called *Angström equation*. This has been generalized to the *Angström–Prescott* (CR6) *equation* by introducing a *daily clearness index*, quantifying all stochastic meteorological influences, as a measure of the atmospheric transparency (Paulescu et al., 2016). Linke (CR3) in 1922 published his *turbidity factor* for the attenuation of the sun's radiation by water vapor and aerosols in the atmosphere. Later Black, Bonython & Prescott (CR12) gave a linear regression relation between solar radiation and the duration of sunshine based on monthly values (Munkhammar & Widen, 2016) and Glover & McCulloch (CR15) improved this by including latitude effects.

Two *engineers*, Hottel & Woertz (CR7), came up with the first serious study on solar energy in 1942: the fundamental relationships given in their classic paper have since then been used for decades to model solar collectors. Hottel & Whillier (CR13) evaluated them concerning the flat-plate solar collector performance (Stanciu, Stanciu, & Paraschiv, 2016) and formulated the *Hottel-Whillier-Bliss equation* on its heat flow and available heat balance, based on considerations of the thermal usability of solar irradiation, coming from Whillier's PhD thesis (CR11) under Hottel's supervision at MIT. This latter work concerned also the relation between radiation on different time scales, showing a close interdependence of the frequency distributions of the so-called clearness index on a monthly, daily, and hourly basis (Vijayakumar, Kummert, Klein, & Beckman, 2005). Because information on sunshine duration was no longer sufficient, he later proceeded to "The determination of hourly values of total solar radiation from daily summations" (CR14) by statistical means, a subject still of great importance for SEM, where still an ever more time-resolved knowledge of solar irradiance is needed.

The *modeling* of solar irradiance components through parameterization of atmospheric phenomena is an equally important area of work in SEM. It was begun in the 1940s by Haurwitz (CR8-CR10), focusing on cloudiness, cloud density, and cloud type (Chowdhury, 1990). (We do not include a publication from the 1948 peak by Penman with NCR=8, i.e. more than CR10, because it is only concerned with evaporation by solar radiation.)

*Time Period 2: 1960 to 1995*

Figure 3 shows the RPYS-CO spectrogram for the marker paper (Liu & Jordan, 1960) after its publication year.

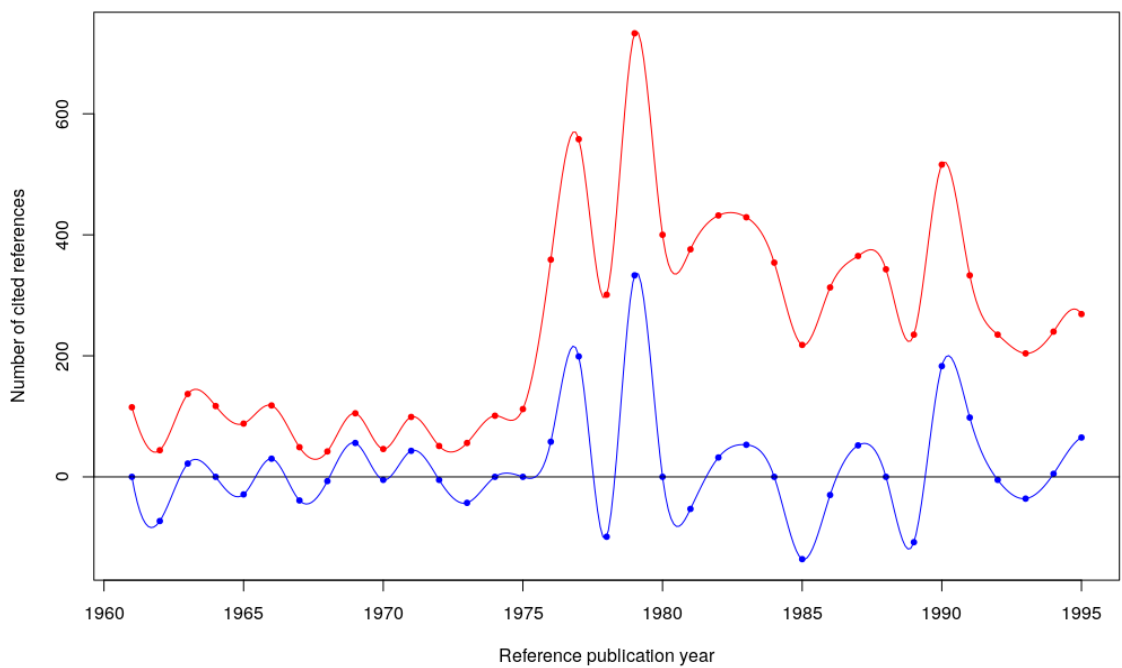

**Figure 3. RPYS-CO graph for (Liu & Jordan, 1960) and period 2 (1960 – 1995) with NCR (red line) and 5-year-median deviation (blue line). For better presentation, the RPY 1960 is excluded.**

Another 10 peaks could be identified from Figure 3 for the following RPYs in the second time period: 1960, 1963/64, 1966, 1969, 1971, 1976/77, 1979/80, 1981-83, 1986-88, 1990/91.

The first two peaks are mainly caused by *engineers*: After the marker paper itself (CR16) the same authors gave generalized curves to predict the "Long-Term Average Performance of Flat-Plate Solar-Energy Collectors", making use of Hottel's and Whillier's methods (CR13) and building upon the knowledge of two parameters only: i) the monthly-average daily total radiation on a horizontal surface and ii) the monthly average day-time ambient temperature (CR17). In 1961, J. K. Page presented "The estimation of monthly mean values of daily total short-wave radiation on vertical and inclined surfaces from sunshine records for latitudes 40N-40S" (CR19) on a "UN Conference on New Sources of Energy" in Rome but the conference proceeding was published in 1964. Much later, he advised advanced publicly funded projects like HELIOSAT-3 (Mueller et al., 2004).

The *meteorologist* F. Kasten (CR22) developed turbidity models as one essential ingredient for radiation model calculations and also functioned as an advisor in later solar energy projects.

Attempts to check and confirm the *diffuse-to-total radiation correlation* by Liu and Jordan (1960) against *measurements* have been undertaken for several locations in the world: Southern Israel by Stanhill (CR20), New Delhi by Choudhury (CR18), and Canada by Ruth and Chant (CR27) and Tuller (CR29).

Cooper (CR23) and Spencer (CR25) are the only representatives in Table 1 of those researchers concerned with *solar geometry*, i.e. sun-earth angle values over time – which is essential for all modeling of radiation. In this respect, our method could only capture these first works, but not later standard works like Michalsky (1988).

Robinson's "Solar Radiation" (CR21) was a meteorological standard publication, but not that much focused as Kondratyev's monograph "Radiation in the Atmosphere" (CR24), whose influence lasted until Iqbal's standard work "Introduction to Solar Radiation" (CR41) from 1983.

In the 1970s and early 80s, one focus of research literature was on *time-resolved diffuse radiation models from the scale of months down to hours*, mostly from the viewpoint of *engineering* like Duffie & Beckman's volumes "Solar Energy Thermal Processes" in 1974 (CR26) and "Solar engineering of thermal processes" in two editions in 1980 (CR37) and 1991 (CR49), but also Hottel (CR28), Orgill & Hollands (CR31), Klein (CR32), and Erbs, Klein & Duffie (CR40). Only Hay (CR30) and Iqbal (CR38) represent *geography* resp. *meteorology*. *Empirical radiation modeling*, in particular with respect to tilted surfaces (e.g., of solar panels), was done by Hay (CR36) and *meteorologists* as Temps & Coulson (CR33) and Klucher (CR35).

At the end of the 1970s, the focus switched also to *stochastic modeling*, outstandingly covered by Collares-Pereira & Rabl (CR34) with their time series analysis and production of the *first synthetic time series*, that were widely used in the community. In the same vein, Bendt presented his "frequency distribution of daily insolation values" (CR39). The time-scale was later even narrowed down to *minute data* by Suehrcke & McCormick (CR44), and Graham, Hollands & Unny (CR45) were able to *simulate daily values* of the clearness index from monthly mean values by using probability distribution functions.

In 1987, Skartveith & Olseth (CR42) presented a *diffuse fraction model*, that became essential part of later works, e.g. in HELIOSAT (Dürr & Zelenka, 2009). CR43, i.e., Perez, Seals, Ineichen, Stewart, and Menicucci (1987), also focused on the diffuse part of total irradiance and accomplished a major improvement in its error-prone computation, in order to "estimate short time step (hourly or less) irradiance on tilted planes" (Perez, et al., 1987), which has received high recognition in the community. (See *Discussion & Conclusion* for considerations to use CR43 as a second marker paper.)

Duffie & Beckman, together with their coauthor Reindl, were mainly responsible for the last high peak, taken into account in our RPYS-CO analysis, in 1990: they *evaluated statistical models for hourly radiation on the tilted surface* (CR47) and could significantly *improve on the time resolution of statistical diffuse radiation models* in CR46, whose abstract we now quote (with our emphases in italics): "The influence of climatic and geometric variables on the *hourly diffuse fraction* has been studied, based on a data set with *22,000 hourly*

*measurements* from five European and North American locations. The goal is to determine if other predictor variables, in addition to the clearness index, *will significantly reduce the standard error* of *Liu- and Jordan-type correlations* (…). Stepwise regression is used to *reduce a set of 28 potential predictor variables* down to *four significant predictors: the clearness index, solar altitude, ambient temperature, and relative humidity*." (Reindl, Beckman, & Duffie, 1990, abstract)

We can in a sense close the circle to our marker paper after exactly 30 years by mentioning CR48, i.e., Perez, Ineichen, Seals, Michalsky, and Stewart (1990), as a successful attempt to *apply diffuse fraction models* to questions of *daylighting in buildings*, transferring irradiance to illumination, and again connecting the fields of *meteorology & radiation* to *energy & engineering*, the two-fold focus of SEM.

**Discussion & Conclusion**

We studied the early history of SEM by applying RPYS-CO to one highly cited and relevant marker paper (Liu & Jordan, 1960), matching the recommendation of a long-term expert, inspected RPYs with peak citation numbers in the corresponding graph and table calculated by the CRExplorer and were able to thereby identify many important papers before and after the marker paper. They give an adequate view of most of the essential contributing streams of research in SEM, as, e.g., measuring, empirical and statistical modeling of direct and diffuse radiation on the horizontal and the tilted plane on time scales from years to minutes. The topics *solar geometry, radiative transfer* calculations through the atmosphere, and *spectrally resolved* treatment of sun light can be identified as underrepresented in our RPYS-CO results. But the latter two in particular gained interest only in later years and should be studied with other marker papers.

It could also be argued that an RPYS-CO on one or few marker papers only produces a bias by favoring a limited number of research groups, maybe even enforced by the effect of self-citations. But in the given case the marker paper is obviously so well chosen as to unearth a diverse set of methodologies and approaches in the world-wide SEM community, coming from the two main domains *meteorology* and *engineering* and covering *measurement*, *modeling* and *evaluation*. (This could be different when we are going to study the satellite based methods, where European and US research groups take slightly different approaches.) Moreover, self-citations of the admittedly repeatedly occurring (co-)authors among the peak papers play no role in our study because of their sporadic appearance.

CR43 (Perez, et al., 1987) had been suggested as a second marker paper by the expert, but as it turned out, the RPYS-CO on both papers only confirms the results for Liu and Jordan (1960) alone, as Figure 4 shows. Some NCR peaks get sharpened, but there are no new ones found. Furthermore, a RPYS-CO analysis on CR43 alone would reveal less peak papers than the RPYS-CO as performed here. An even stronger confirmation results from another RPYS-CO conducted for the two top most cited papers in our list of CRs, i.e., CR34 (Collares-Pereira & Rabl, 1979) with 238 citations and CR40 (Erbs, Klein, & Duffie, 1982) with 208 citations: *all* peaks and peak papers get reproduced, except from one in 1948 (CR10)!

Moreover, none of the potential candidates in our list of CRs got nearly as many citations as Liu and Jordan (1960) in the whole Web of Science. These facts corroborate the careful choice of the marker paper and the stability of the method's outcomes.

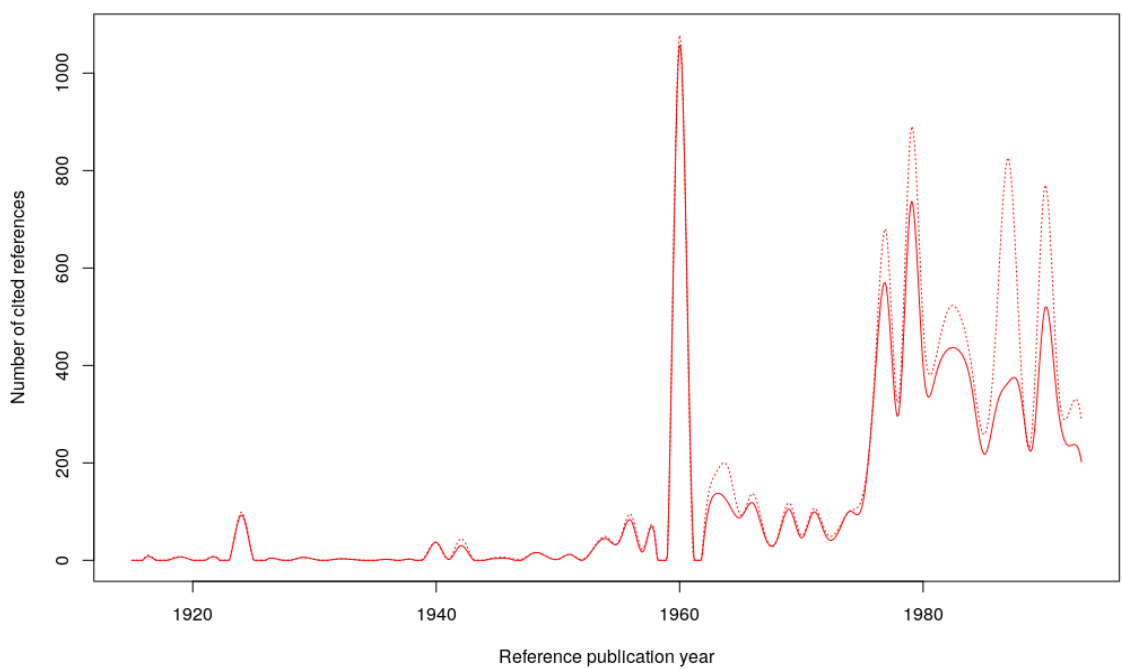

**Figure 4. Comparison of the RPYS-CO for the marker paper Liu and Jordan (1960) (solid line) and the RPYS-CO for the marker paper plus the potential second marker paper Perez, et al. (1987) (dotted line).**

In total, the outcome of our study meets our expectations: All relevant historical roots of SEM research were disclosed by our RPYS-CO analysis. Therefore, we recommend RPYS-CO for similar investigations by researchers to get more insight in the development of their field of work or even as a tool for studies in the history of science.

## Acknowledgments

We want to thank Detlev Heinemann for very useful discussions on the "early story of Solar Energy Meteorology" and comments regarding an earlier version of this manuscript. We also thank Lutz Bornmann for helpful discussions.